\documentclass[letterpaper,english,journal=ascefj,manuscript=letter]{achemso}

\usepackage{textcomp}

\usepackage[T1]{fontenc}
\usepackage{units}
\usepackage{amsmath}
\usepackage{amssymb}
\usepackage{graphicx}
\usepackage{wasysym}
\usepackage{babel}

\SectionsOn
\setkeys{Gin}{width=7cm}

\author{A. Daniel McCartt}
\affiliation{Center for Accelerator Mass Spectrometry, Lawrence Livermore National
Laboratory, ~~7000 East Avenue, Livermore, CA 94550, USA}
\email{mccartt1@llnl.gov}
\author{Jun Jiang}
\affiliation{Center for Accelerator Mass Spectrometry, Lawrence Livermore National
Laboratory, ~~7000 East Avenue, Livermore, CA 94550, USA}

\title{Room-temperature optical detection of $^{14}$CO$_{2}$ below the
natural abundance with two-color cavity ring-down spectroscopy}

\begin{document}
\begin{abstract}
Radiocarbon\textquoteright s natural production, radiative decay,
and isotopic rarity make it a unique tool to probe carbonaceous systems
in the life and earth sciences. However, the difficulty of current
radiocarbon ($^{14}$C) detection methods limits scientific adoption.
Here, two-color cavity ring-down spectroscopy detects $^{14}$CO$_{2}$
in room-temperature samples with an accuracy of one-tenth the natural
abundance in 3 minutes. The intra-cavity pump-probe measurement uses
two cavity-enhanced lasers to cancel out cavity ring-down rate fluctuations
and strong one-photon absorption interference (>10,000 1/s) from hot-band
transitions of CO$_{2}$ isotopologues. Selective, room-temperature
detection of small $^{14}$CO$_{2}$ absorption signals (<1 1/s) reduces
the technical and operational burdens for cavity-enhanced measurements
of radiocarbon, which can benefit a wide range of applications like
biomedical research and field-detection of combusted fossil fuels.
\vfill     {\centering LLNL-JRNL-831265\par}
\end{abstract}

The history of a measured sample can be revealed by the presence of
trace species within its constituents. Species with more unique provenance,
rare abundance, or association with a particular phenomenon offer
more unambiguous distinctions with their detection. One of the most
famous trace species is radiocarbon ($^{14}$C). It is a rare isotope
of carbon (abundance $1.2\times10^{-12}$ $\textrm{\ensuremath{\nicefrac{^{14}\textrm{{C}}}{\textrm{{C}}}}}$),
which is produced in the atmosphere by the interaction of cosmogenic
neutrons with nitrogen  and radiatively decays with a half-life of
\textasciitilde 5700 years.\cite{Libby1946,osti_380330} These unique
properties of radiocarbon can be leveraged for a wide variety of scientific
applications (\emph{e.g.}, radiocarbon dating, isotope tracer studies,
cosmogenic nuclide analysis, etc.), but the rarity of $^{14}$C makes
measurements challenging.\cite{Bennett1977,Turteltaub2000,Heaton2021}
In the 1970s, nuclear physicist employed accelerator mass spectrometry
(AMS) to increase the sensitivity of $^{14}$C detection. These room
sized instruments utilized megavolt accelerators to separate $^{14}$C
from its main interfering isobars ($^{14}$N and $^{13}$CH) and achieved
unprecedented sensitivity (\emph{i.e.}, $10^{-16}$ $\textrm{\ensuremath{\nicefrac{^{14}\textrm{{C}}}{\textrm{{C}}}}}$).\cite{Nelson1977,Bennett1977} 

While $^{14}$C detection with AMS was initially developed for radiocarbon
dating, $^{14}$C's isotopic rarity in concert with AMS's sensitivity
can track small quantities of $^{14}$C-labeled carbonaceous species
through chemical systems. For example, the safety and efficacy of
novel drugs can be tested in humans by administering $^{14}$C-labeled
microdoses. These microdoses are small enough that they are not therapeutic,
but $^{14}$C-tagged metabolites produced by the body can still be
detected.\cite{Turteltaub2000} 

The natural production and radiative decay properties of $^{14}$C
provide opportunities to monitor the carbon cycle and fossil-fuel
emissions.\cite{Heaton2021,Levin2003,Turnbull2009} Atmospheric CO$_{2}$
is in constant flux with terrestrial and marine ecosystems at rates
20 times greater than human emissions. This makes the attribution
of atmospheric CO$_{2}$ measurements to emission point sources of
combusted fossil fuels difficult and inaccurate. By measuring the
dilution of atmospheric $^{14}$CO$_{2}$ with combusted fossil fuels
depleted of $^{14}$C, anthropogenic emissions can be monitored and
differentiated from natural CO$_{2}$ fluxes.\cite{Levin2003,Turnbull2009}
These atmospheric $^{14}$CO$_{2}$ tracer measurements can verify
``bottom-up'' accounting estimates of fossil-fuel emissions and
create a framework of accountability and trust for emission reduction
agreements.\cite{Lauvaux2020,Miller2020,Gurney2021} While scientists
continue to find new applications using $^{14}$C, the size, cost,
and complexity of AMS limit these endeavors, particularly for high-throughput
and fieldwork measurements.\cite{DellOrco2021,Gurney2021,Kratochwil2018,Miller2020}
This has spurred the development of alternative and more accessible
means of $^{14}$C detection.

Cavity ring-down (CRD) spectroscopy has emerged as a viable $^{14}$C
detection technique.\cite{Genoud:15,McCartt2015,Galli2011b,Galli2016,McCartt2016,Fleisher2017,Sonnenschein2018}
It utilizes strong anti-symetric-stretch band ($\nu_{3}$) transitions
of $^{14}$CO$_{2}$ in the mid-IR and has demonstrated parts-per-quadrillion
(ppq) precisions ($10^{-15}\textrm{\:\ensuremath{\nicefrac{^{14}\textrm{{C}}}{\textrm{{C}}}}}$),
which are well below the natural abundance (\emph{i.e.}, $1.2\times10^{-12}$
$\textrm{\ensuremath{\nicefrac{^{14}\textrm{{C}}}{\textrm{{C}}}}}$).\cite{Galli2011b,Galli2016}
A high-finesse, optical cavity that is centimeters long can provide
gas-laser interaction path lengths equivalent to kilometers. This
cavity-enhanced path length increases the total gas absorption sensitivity,
but this enhancement is not selective to the target analyte alone.
Drifts in the cavity base loss and absorption interference from other
species reduce the accuracy of traditional, cavity-enhanced techniques.
Spurious reflections and external etalons coupled to the cavity cause
frequency and time dependent undulations in the cavity base loss.\cite{Huang2010}
Drifts in these unwanted signals compromise traditional CRD trace
gas measurements. Near the $^{14}$CO$_{2}$ $\nu_{3}$ band in the
Mid-IR, absorption interference is strong from CO$_{2}$-isotopologue
hot-band transitions and other molecular species such as N$_{2}$O.
Previous CRD measurements of $^{14}$CO$_{2}$ with accuracy below
the natural abundance mitigated against this interference by cooling
the test gas.\cite{Galli2011b,Galli2016,McCartt2016,Fleisher2017}
However, the most accurate measurements cooled with two-stage refrigeration
units, dry-ice baths, or Stirling engines, which negated the reduced
size and portability benefits of laser based techniques. Even after
cooling, the extremely weak $^{14}$CO$_{2}$ signal had to be extracted
from the dense interference (100 times smaller at natural abundance,
20 torr, and -20 \textdegree C). This required laser wavelength scans
and spectroscopic line-shape fitting of multiple overlapping features.
While CRD has demonstrated $^{14}$CO$_{2}$ sensitivities below the
natural abundance which can be of service to multiple scientific fields,
the difficulty of these measurements has largely confined them to
laser spectroscopy laboratories. 

Here, we present the first two-color, cavity ring-down spectroscopy
(2C-CRDS) measurements of room-temperature $^{14}$CO$_{2}$ samples.
This recently developed intra-cavity pump-probe technique uses an
additional cavity-enhanced pump laser to selectively extract the signal
of interest and cancel out instrument drift and unwanted background
absorption interference.\cite{Jiang2021a} Gas cooling and cavity-base-loss
stabilization are no longer required for $^{14}$CO$_{2}$ sensitivity.
The burden of signal extraction is transferred from spectroscopic
analysis to an intrinsic 2C-CRDS capability, and because of its noise
cancelation, experimental conditions that would have been prohibitively
unstable are now a possibility (\emph{e.g.,} flow-through and field
measurements). The unprocessed 2C-CRDS signal accurately detects $^{14}$CO$_{2}$
concentrations with one tenth of the natural abundance. This is achieved
in three minutes, at room temperature, and without spectral-fitting
compensation for interfering species or cavity-base-loss variations. 

These results exceed requirements for most biological and biomedical
$^{14}$C-tracer experiments. However, as is evident by the 2C-CRDS
precision with the cavity under vacuum (equivalent to $7\permil$
of the natural abundance, see supporting information), there is room
for improving the 2C-CRDS measurement accuracy of $^{14}$CO$_{2}$.
We determine precision from Allan deviation analysis that provides
information about the stability of the 2C-CRDS signal during a single
measurement. Accuracy is obtained from the mean absolute error of
a linear fit to multiple sample types measured over a period of weeks.
Factors that are contributing to the discrepancy between the demonstrated
accuracy (91 ppq) and precision (33 ppq) are discussed, and spectroscopic
strategies for overcoming this difference are presented. Improvements
that bring the accuracy into line with the demonstrated precision
would allow 2C-CRDS to address more demanding applications such as
field measurements of fossil fuel emissions. This could transform
existing \textquotedblleft top-down\textquotedblright{} atmospheric
CO$_{2}$ monitoring systems to constrain \textquotedblleft bottom-up\textquotedblright{}
estimates of fossil fuel emissions, provide stakeholders with time
and space resolved emissions data of both natural and fossil fuel
derived CO$_{2}$, and create a platform of accountability and trust
for national emission reduction commitments.

\section*{Experimental}

\begin{figure}[!b]
\centering\includegraphics[width=0.9\textwidth]{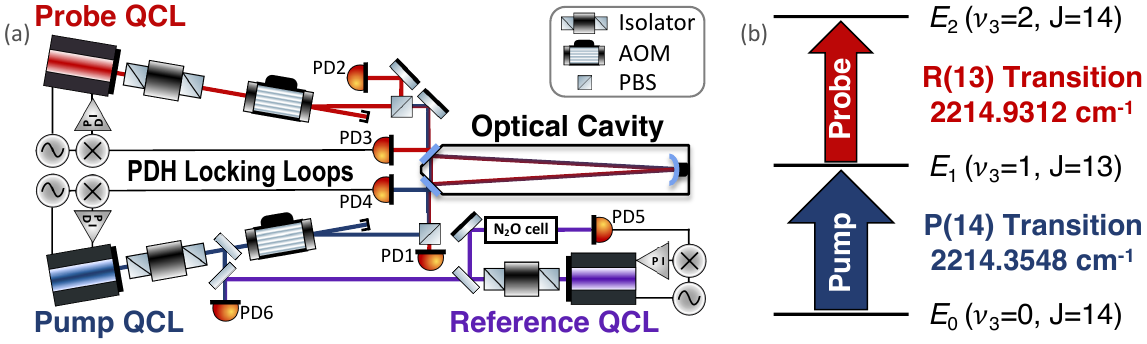}\caption{Summary of the experimental methods used for this study. A schematic
of our 2C-CRDS experimental setup shown in panel (a), and the diagram
of the ladder-type $^{14}$CO$_{2}$ P(14)-R(13) pump-probe scheme
is presented in panel (b). Both the pump (blue) and probe (red) QCLs
are locked to the high-finesse optical cavity using PDH signals recorded
at photo-detectors PD3 and PD4. The pump power is monitored with PD2
and ring-downs are recorded with PD1. Finally an additional reference
QCL is locked to the $\nu_{3}$ 1 \textendash{} 0, R(16) transition
of $^{15}$N$^{14}$N$^{16}$O . The beatnote between this laser and
the pump recorded at PD6 provides a frequency reference for the 2C-CRDS
experiments.  PBS and AOM stand for polarization beam splitter and
acousto-optic modulator. \label{fig:Schematic-of-2C-CRDS}}
\end{figure}

 Figure \ref{fig:Schematic-of-2C-CRDS} shows a schematic with the
primary components of our 2C-CRDS setup and a three level diagram
of the $^{14}$CO$_{2}$ pump-probe scheme used in this paper. Two
quantum cascade lasers (QCL) are injected into a high-finesse optical
cavity in a counter propagating beam configuration. S- and p-polarization
modes of the three-mirror, traveling-wave cavity share a common beam
path but have different resonant cavity frequencies.\cite{Saraf2007}
 Using the Pound-Drever-Hall (PDH) technique, the pump laser is locked
to a lower-finesse p mode $\left(\mathcal{F_{\textrm{{p}}}=}5280\right)$
close to the $^{14}$CO$_{2}$, $\nu_{3}$-band 1 \textendash{} 0,
P(14) transition.\cite{Drever1983} The probe laser is locked to a
high-finesse s mode $\left(\mathcal{F_{\textrm{{s}}}=}67700\right)$
near the $\nu_{3}$-band 2 \textendash{} 1, R(13) transition. Ring-down
measurements are initiated by periodically interrupting the probe
laser's lock to the cavity. In between each ring-down event, the
intra-cavity pump power is applied or removed by cycling the pump
laser's lock on and off. The difference between these ``pump-on''
and ``pump-off'' ring-downs provide a net, two-color signal sensitive
to the pumped $\nu_{3}=1,\:J=13$ population and compensated for instrument
drift and one-color spectroscopic interference. Further details on
the 2C-CRDS technique can be found in our previous paper on N$_2$O.\cite{Jiang2021a} 

The frequency spacings of the cavity resonant modes and the upper
and lower molecular transitions influence the appearance of the intra-cavity
two-color spectra.\cite{Jiang2021a} The lasers are locked to a resonant
mode of the cavity p- or s-polarization transmission ``combs'' and
are simultaneously tuned when the cavity length is changed. Modes
of a given polarization are separated by the cavity free spectral
range ($FSR$), and for a three-mirror cavity, the s and p ``combs''
are interleaved with a spacing of approximately half the $FSR$.\cite{Saraf2007}
These experimental constraints of the cavity discretize the frequency-separation
selection between the pump and probe lasers and when combined with
the transition frequencies of the analyte, dictate the appearance
and position of the intra-cavity two-color spectroscopic features.

Because of the high, intra-cavity, pump-laser power (17 W) and the
counter-propagating beam configuration, 2C-CRDS measurements can exhibit
several qualitatively different spectroscopic features. For $^{14}$CO$_{2}$,
coherent, step-wise, and Autler-Townes-splitting features are observed
with this experimental setup. The coherent features occur when the
sum of the pump and probe laser frequencies is equal to the sum of
the upper and lower molecular transition frequencies. These peaks
are effectively Doppler-free as the near-identical frequencies of
the counter propagating beams cancel out Doppler shifts from longitudinal
velocities. The step-wise peaks are Doppler broadened and, unlike
the coherent features, populate the intermediate $\nu_{3}=1$ level
during a two-step absorption process. Regardless of the pump detuning,
step-wise peaks always occur at near-zero probe detuning. Finally,
for features where both the pump and probe are near resonance, an
Autler-Townes-type splitting is observed.\cite{Autler1955}

\section*{Results and Discussion}

\begin{figure}[t]
\centering\includegraphics[width=10cm]{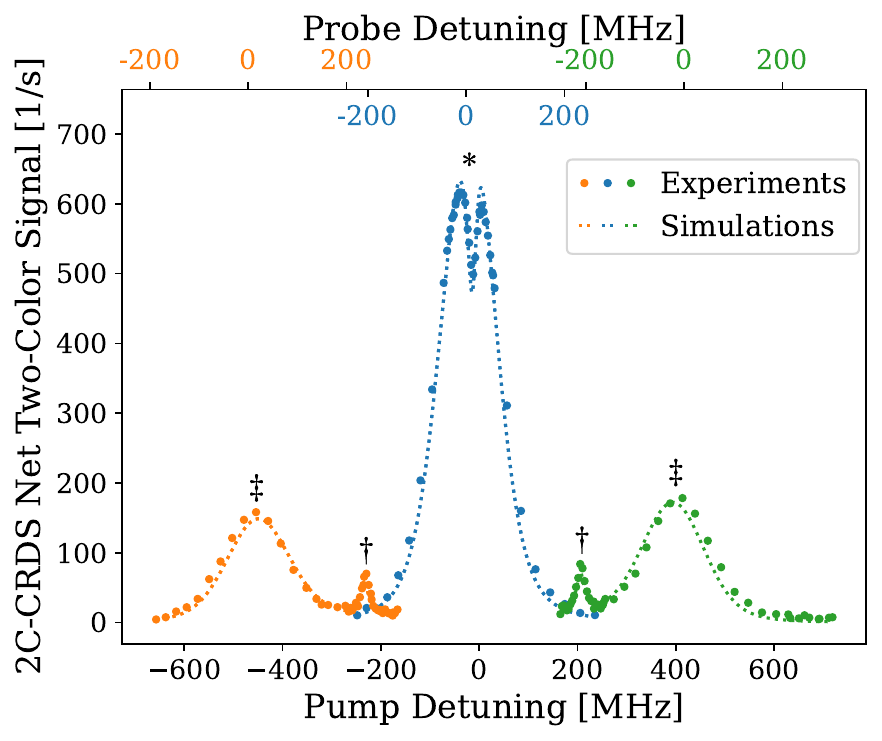}

\caption{2C-CRDS spectra of $^{14}$CO$_{2}$ from combusted glucose with 375
times the natural abundance of $^{14}$C. Measurements were taken
at room temperature and 4 torr. Note that the spectra are plotted
against pump detuning on the lower axis, and the shifted probe detunings
are shown on the color-coded upper axes for each of the three pump-probe
spacings. Features are annotated with the following symbols: coherent
$\dagger$, step-wise $\ddagger$, Autler-Townes-type {*}.  \label{fig:2C-CRDS-Spectrum}}
\end{figure}
Figure \ref{fig:2C-CRDS-Spectrum} shows 2C-CRDS spectra of the $^{14}$CO$_{2}$
1 \textendash{} 0, P(14) and 2 \textendash{} 1, R(13), intra-cavity,
two-color transitions for a room-temperature, 4-torr, carbon-dioxide
sample with 375 times the natural abundance of $^{14}$C. Three spectra
from experiments where the pump and probe lasers were locked to different
sets of cavity modes are plotted alongside simulations. The $^{14}$CO$_{2}$,
2C-CRDS density matrix simulations are analogous to what was presented
for our N$_2$O study.\cite{Jiang2021a} Step-wise and coherent, two-color
peaks were observed for experiments where the probe is locked to the
mode on resonance with the $^{14}$CO$_{2}$, 2 \textendash{} 1, R(13)
probe transition and the pump was locked to a cavity mode detuned
by one $FSR$ (443 MHz or \textasciitilde 7 times the Doppler half
width) from the $^{14}$CO$_{2}$, 1 \textendash{} 0, P(14) pump transition
center frequency. When combined with a high-resolution measurement
of the 1 \textendash{} 0, P(14) transition,\cite{Galli2011} the
Doppler-free, coherent features provide an accurate quantification
of the previously-unmeasured 2 \textendash{} 1, R(13) transition
wavenumber at 2213.9319(3) cm$^{-1}$ (accuracy limited by the beat-note
frequency reference). This is 0.017 and 0.014 cm$^{-1}$ greater than
\emph{ab initio} calculations for the 2 \textendash{} 1, R(13) transition.\cite{Zak2017,Huang2017}
For this experiment's cavity geometry, the $^{14}$CO$_{2}$, 1 \textendash{}
0, P(14) and 2 \textendash{} 1, R(13), transition-pair frequencies
are separated by nearly an exact odd integer multiple of the discretized
p- and s-mode frequency spacing (\emph{i.e.}, \textasciitilde$57\times\nicefrac{FSR}{2}$ ).
This allows the pump and probe lasers to be simultaneously tuned on
resonance with their respective transitions. In this case with the
lower pump level of the ladder-type system driven by a large effective
Rabi frequency (\textasciitilde 55MHz) and near identical pump and
probe laser frequencies, the 2C-CRDS spectroscopic feature exhibits
Autler-Townes-type splitting, which is well reproduced by the density
matrix simulation.\cite{Salomaa1976,Lee2000} 

\subsection*{Measurements of $^{14}$C standards}

\begin{figure}[h]
\centering{}\centering\includegraphics[width=1\textwidth]{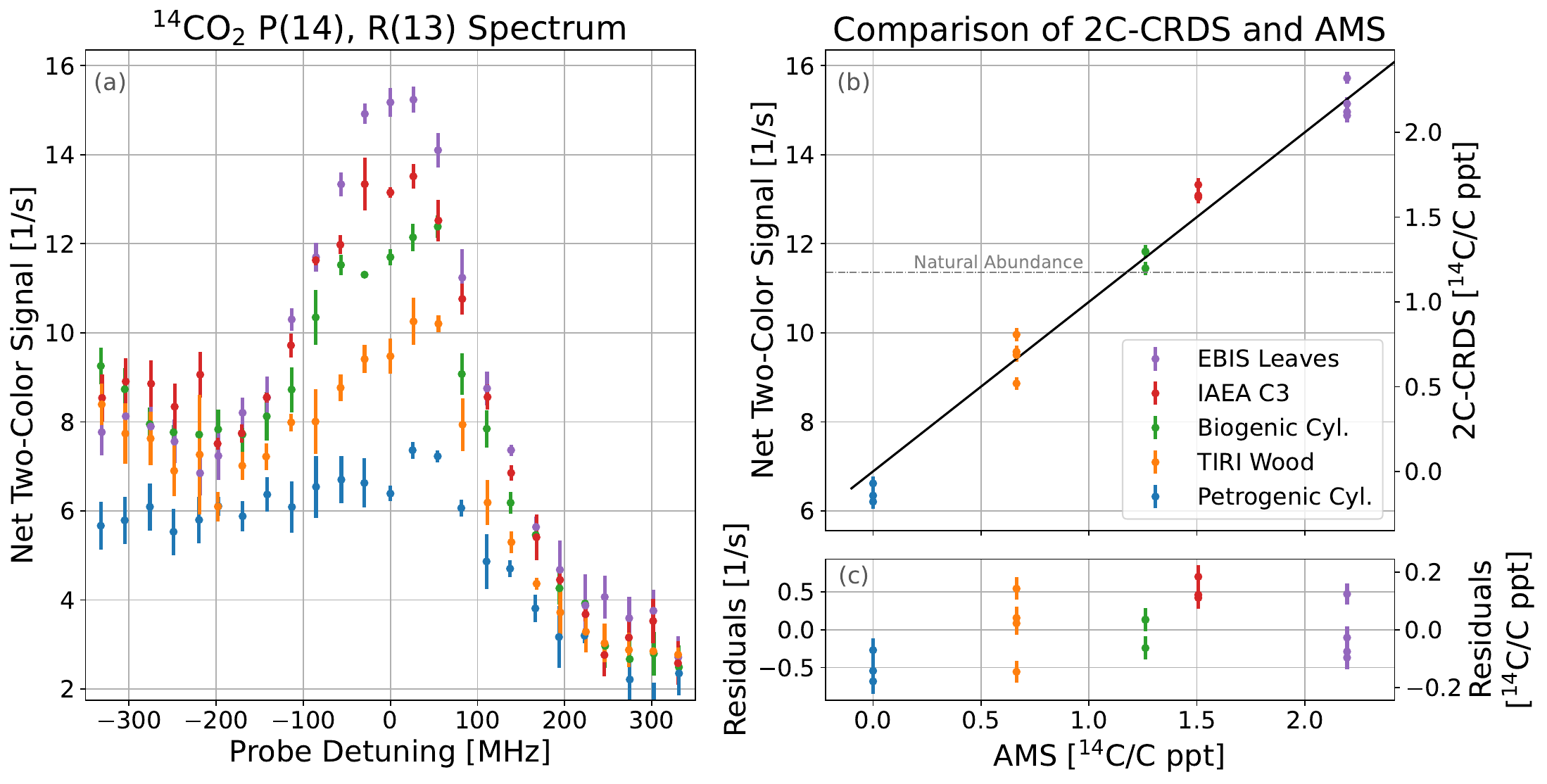}\caption{2C-CRDS measurements of samples containing zero to two times the natural
abundance of $^{14}$C. Panel (a) shows 2C-CRDS spectra averaged for
a given sample type, and error bars represent the $2-\sigma$ standard
deviation of these groupings. Panel (b) shows the comparison of 2C-CRDS
measurements to AMS. The error bars are the 95\% confidence interval
for a single two-color measurement with the probe on resonance. A
linear fit of the comparison is used to infer equivalent carbon-14
concentrations (axes on right), and the slope is constrained by a
combusted glucose sample which contains 30 times the natural abundance
of carbon-14 (not shown). The net two-color signal for a sample containing
the natural abundance of $^{14}$C is indicated by a dash-dot style
line. Panel (c) displays the residuals of the fit. The legend in panel
(b) is for all three panels.   \label{fig:AMS-comparison}}
\end{figure}

2C-CRDS's $^{14}$C detection capabilities were characterized with
CO$_2$ samples from commercial sources and combusted ``$^{14}$C
standards.'' Spectra were taken at room temperature and 20 torr (Figure
\ref{fig:AMS-comparison}a), and the $^{14}$C concentrations ranged
from petrogenic ($\sim$zero $\textrm{\ensuremath{\nicefrac{^{14}\textrm{{C}}}{\textrm{{C}}}}}$)
to approximately double the natural abundance (\emph{i.e.}, 1.2 parts
per trillion (ppt) $\textrm{\ensuremath{\nicefrac{^{14}\textrm{{C}}}{\textrm{{C}}}}}$).
At 20 torr, the spectroscopic feature complexity is greatly reduced
with the step-wise resonances dominating and the Autler-Townes-type
effect minimized as a result of collisional broadening. 2C-CRDS resolves
$^{14}$CO$_{2}$ concentration differences that are fractions of the
natural abundance in samples sourced from: petrogenic fuel (Petrogenic
Cyl.), a 5240-year-old tree (TIRI Wood, Pinus sylvestris), 70-year-old
cellulose (IAEA C3), contemporary corn (Biogenic Cyl.), and $^{14}$C
elevated leaves collected near a medical waste facility (EBIS Leaves).\cite{Scott2003,Rozanski1992,Trumbore2002,Swanston2005,Hanson2005}
To demonstrate the detection selectivity of the 2C-CRDS method, the
unprocessed two-color signal with the probe on resonance is compared
with duplicate sample analysis by AMS (Figure \ref{fig:AMS-comparison}b).
Residuals from a linear fit of this comparison have a mean absolute
error of 91 ppq\textendash an accuracy that is better than 8\% of
the natural $^{14}$C abundance. Accuracy can be improved to 5\% by
averaging the results of each sample type and effectively increasing
the averaging time. However, the rate of error reduction from averaging
decreases after 3 minutes (see Figure S2). For the P(14)-R(13) pump-probe
scheme and this set of samples, an elevated and varying background
in the two-color signal (Figure \ref{fig:AMS-comparison}a) is likely
affecting averaging performance and this should be compensated or
removed to increase the accuracy of 2C-CRDS $^{14}$CO$_{2}$ detection.

\subsection*{Collisionally induced background}

\begin{figure}[h]
\centering\includegraphics[width=10cm]{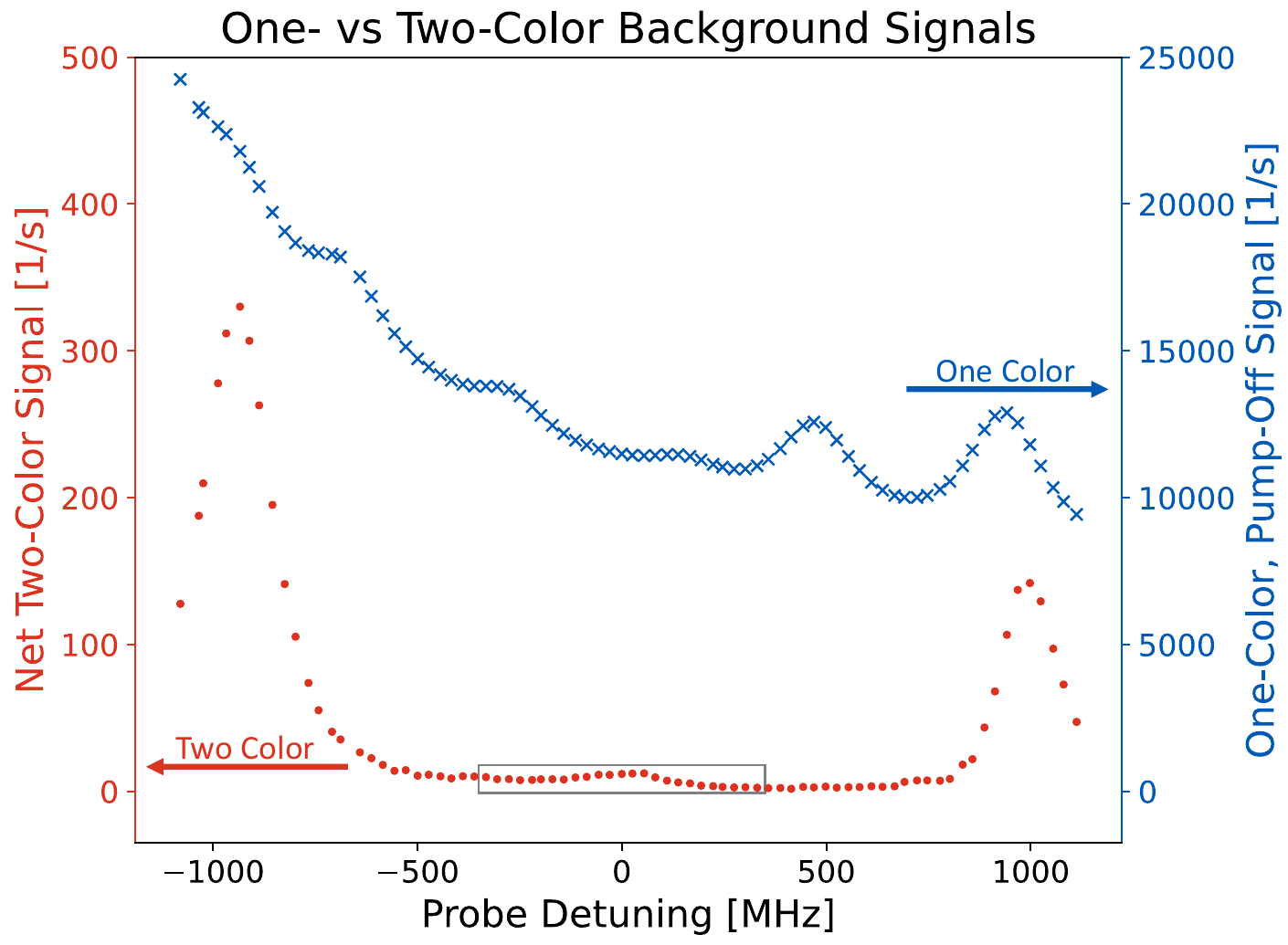}

\caption{One- vs two-color background signals for the biogenic cylinder sample
in the region surrounding the 2C-CRDS, $^{14}$CO$_{2}$, 1 \textendash{}
0 P(14), 2 \textendash{} 1 R(13), two-color resonance. The net, two-color
signal axis is on the left and the one-color signal axis is on the
right. The grey rectangle surrounding some of two-color points delineates
the area corresponding to the spectra shown for the ``$^{14}$C-standards''
(Figure \ref{fig:AMS-comparison}a). While only probe detuning is
indicated, the pump is scanned over an equivalent frequency range.
\label{fig:One--and-two-color-background}}
\end{figure}

For the ``$^{14}$C-standard'' samples measured here, the one-color
spectra are dominated by absorption interference from hot-band transitions
of CO$_{2}$ isotopologues. 2C-CRDS reduces this one-color interference
by 3 orders of magnitude (Figure \ref{fig:One--and-two-color-background}).
However, additional two-color features are also present, which are
contributing to the background at the $^{14}$CO$_{2}$, 2 \textendash 1,
R(13) probe frequency. At zero probe detuning, this background signal
is similar in magnitude to the natural abundance $^{14}$CO$_{2}$
signal (Figure \ref{fig:AMS-comparison}a). Furthermore, off resonance
(\emph{e.g.}, -300 MHz probe detuning) there are statistically significant
differences between samples in the background signal. 

Like traditional pump-probe techniques, 2C-CRDS filters interfering
transitions for species that are excited at both laser frequencies.
This includes conventional quantum-state-linked transition pairs (\emph{e.g.},
$^{14}$CO$_{2}$ 1 \textendash{} 0, P(14) and 2 \textendash{} 1, R(13))
but also collisionally induced two-color signals. The intense intra-cavity
pump power can excite species with relatively far detuned transitions
to vibrationally excited states. Populations are then redistributed
to other vibrational and rotational states via collisions. For abundant
species, even weak population transfer pathways from highly excited
vibrational levels (\textasciitilde 5000 $\textrm{cm}^{-1}$ internal
energy) can create signals significant to trace gas detection. Both
\emph{positive} and \emph{negative} two-signals can occur depending
on whether the pump-perturbed redistribution \emph{increases} or \emph{decreases
}the probed state population. Inaccuracies from these unwanted two-color
signals are exacerbated by their increased pump power dependence.
Unlike the well saturated $^{14}$CO$_{2}$ P(14) transition which
has near zero pump detuning, the collisionally induced signals are
likely pumped by relatively far detuned transitions and would be less
saturated. Understanding the origin and variations of these background
signals is important for maximizing the accuracy of 2C-CRDS detection,
and we intend to conduct a more comprehensive study of the system
dynamics. 

2C-CRDS was designed to cancel out cavity ring-down base-loss variations
and spectroscopic interference, but the effect of some experimental
parameters on the two-color dynamic system needs further characterization.
For example, Allan deviation analysis of vacuum 2C-CRDS measurements
has optimal averaging and a resultant precision of 0.03 $[\nicefrac{1}{s}]$
in 2.3 hours (supplementary text). When CO$_{2}$ is introduced into
the cavity precision is only improved with averaging times up to
10 minutes ($0.127\:[\nicefrac{1}{s}]$,33 ppq $\textrm{\ensuremath{\nicefrac{^{14}\textrm{{C}}}{\textrm{{C}}}}}$).
There is a clear correlation between the two-color signal and the
intra-cavity pump power, and correcting for this variation with an
empirically-determined, power-dependent model increases the time at
which the minimum of the Allan deviation occurs. These pump power
variation levels will not have an appreciable affect on the well saturated
two-color $^{14}$CO$_{2}$ signal (Figure S3), but they are likely
causing significant variation in the two color background. The transmitted
pump power drifted as the QCL thermally settled after power-up and
with day-to-day changes in experimental conditions. Sample to sample
pump power variation likely accounts for some of the two-color spread
in Figure \ref{fig:AMS-comparison} and the discrepancy between the
accuracy (91 ppq) and precision (33 ppq).

The selectivity of the uncompensated, room-temperature, two-color
signal accurately detects a tenth of the natural $^{14}$C abundance,
which exceeds the requirements for nearly all $^{14}$C-tracer applications
in biology and medicine. 2C-CRDS development was funded to detect
$^{14}$C in biological samples separated by liquid chromatography
(LC). Flow-through measurements have already been demonstrated in
our earlier work.\cite{Jiang2021a} This capability combined with
its single frequency measurement selectivity should enable 2C-CRDS
to make laser-based, in-line-flow-through measurements of $^{14}$C
in LC effluent. 

To push the limits of 2C-CRDS $^{14}$C detection, the small, two-color
background variations must be addressed. While a model could be developed
to compensate for variations in the P(14)-R(13) two-color background,
there is likely another $^{14}$CO$_{2}$ two-color line pair with
less interference. In contrast to one-color measurements of $^{14}$CO$_{2}$
that have a consensus transition ($\nu_{3}$ 1\textendash 0, P(20))
and rely on further mitigation of determined interference to increase
sensitivity (test-gas cooling, spectroscopic line shape fitting and
compensation, etc.), 2C-CRDS still has many unmeasured two-color $^{14}$CO$_{2}$
line pairs that could provide better opportunities for $^{14}$C detection.
Within the tuning range of the QCL's used in this study, P(14)-R(13)
was the quantum-state-linked pair with the smallest two-color background,
but collisionally induced signals can also be used for $^{14}$CO$_{2}$
detection.  A preliminary investigation using the collisionally-assisted
1\textendash 0, P(20) and 2\textendash 1, R(13) pump-probe scheme
yields a reduced two-color background, and the $^{14}$CO$_{2}$ signal
is of similar strength to the linked two-color scheme used in this
work due to facile rotational population relaxation within the $\nu_{3}=1$
manifold. The inclusion of collisionally-assisted pump-probe schemes
significantly increases $^{14}$CO$_{2}$ two-color line-pair options
and the likelihood of a set with reduced and stable background. 
After identifying the ideal line pair and constructing a model to
compensate for any remaining two-color background signal, we are confident
the accuracy of 2C-CRDS $^{14}$CO$_{2}$ detection can be brought
into line with the demonstrated precision.

While $^{14}$CO$_{2}$ is a particularly demanding case study, 2C-CRDS
can be applied to other species for trace-gas detection and quantum-state-resolved
molecular spectroscopy. High-energy transitions, similar to those
causing the interfering two-color background, are of interest to astrophysical
observations. 2C-CRDS's Doppler-free coherent resonances can accurately
determine high-energy transition frequencies which are difficult to
measure through thermal excitation and can be used to investigate
the atmospheric composition of exoplanets.\cite{Foltynowicz2021,Showman2008,Guilluy2019,Hu2020}

\section*{Conclusion}

2C-CRDS integrates the selectivity of pump-probe techniques with the
sensitivity of cavity-enhanced detection. The resultant two-color
spectrum focuses the sensitivity of the cavity on the species of interest
and cancels out drift typical of cavity-enhanced instrumentation.
These qualities are ideal for trace-gas measurement. Furthermore,
the pump-probe line-pair selection of 2C-CRDS affords experimental
flexibility. 2C-CRDS can be tailored to compensate for interfering
species prevalent in a given application and designed to maximize
noise cancelation for field work. Careful study of the $^{14}$CO$_{2}$
two-color line-pair options and associated background signals is needed
to realize the full $^{14}$C 2C-CRDS detection potential, but the
technique shows promise for atmospheric monitoring of anthropogenic
CO$_{2}$ emissions. $^{14}$CO$_{2}$\textquoteright s utility for
tracking fossil fuel emissions has been demonstrated on scales ranging
from individual point sources to continents, and can provide an objective
metric for coalitions of nations and municipalities that have agreed
to reduce green-house-gas emissions.\cite{Gurney2012,Basu2020} 2C-CRDS's
unique combination of sensitivity and field ability offer a logistically
feasible and affordable means of consistently monitoring combusted
fossil-fuel emissions with $^{14}$CO$_{2}$. 
\begin{acknowledgement}
The authors thank Professor Kevin K. Lehmann (UVA) and Dr. Davide
Mazzotti (CNR-INO) for their thoughtful comments on the manuscript,
and Ted Ognibene, Kari Finstad, Kurt Haack, Alexandra Hedgpeth, Caroline
Stitt, and Esther Ubick (LLNL) for their assistance with the various
samples.

Research reported in this publication was supported by the National
Institute of General Medical Sciences of the National Institutes of
Health (Award No. R01GM127573). The content is solely the responsibility
of the authors and does not necessarily represent the official views
of the National Institutes of Health. This work was performed, in
part, at the National User Resource for Biological Accelerator Mass
Spectrometry, which is operated at the LLNL under the auspices of
the U.S. Department of Energy (Contract No. DE-AC52-07NA27344).The
user resource was supported by the National Institutes of Health,
National Institute of General Medical Sciences (Grant No. R24GM137748).
\end{acknowledgement}
\begin{suppinfo}
Additional experimental details, materials, and methods, supplementary
text, including Figures S1, S2, and S3 (PDF)
\end{suppinfo}

\bibliography{citations_fixSubp_v4}

% \begin{tocentry}
% \includegraphics[width=9cm]{\string"./TOC_v3\string".pdf}
% \end{tocentry}
\end{document}

\baselineskip24pt

This PDF file includes:

\qquad{}Materials and Methods 

\qquad{}Supplementary Text 

\qquad{}Figures S1, S2, and S3

\vfill     {\centering LLNL-JRNL-831265\par}    \newpage

\section*{Materials and Methods}

\subsection*{$^{14}$C Materials}

The samples containing 30 and 375 times the natural abundance of $^{14}$C
are from dilutions of a combusted glucose mixture (Sigma G-5020 and
Mallinckrodt 4912). The ``Petrogenic Cylinder'' sample is instrument
grade CO$_2$ sourced from petroleum feedstock (Praxair). The ``Biogenic
Cylinder'' sample is also instrument grade CO$_2$ but sourced from
ethanol-based production using contemporary corn (\textasciitilde 2015)
as the feedstock (Airgas). The ``TIRI Wood'' sample is ``Belfast
Pine, Sample B'' from the Third International Radiocarbon Intercomparison
with designator Q7780.\cite{Scott2003} ``IAEA C3'' is cellulose
produced in 1989 from \textasciitilde 40 year old trees and used
in the ``The IAEA $^{14}$C Intercomparison Exercise 1990''.\cite{Rozanski1992}
``EBIS Leaves'' are from the ``Enriched Background Isotope Study''
and were collected near a medical waste facility that was emitting
elevated levels of $^{14}$C.\cite{Trumbore2002,Swanston2005,Hanson2005}

\subsection*{Sample Preparation}

Duplicate samples containing up to 5 mg of carbon were aliquoted and
dried under vacuum. These samples and approximately 150 mg of copper
oxide were sealed in a quartz tubes using an acetylene torch. The
tubes were then combusted at 900$\degree$C for 2 hours. Following
combustion, AMS samples were graphitized and 2C-CRDS samples were
transferred to the spectrometer front-end for purification. For 2C-CRDS
sample purification, the quartz tubes were cracked under vacuum inside
a bellows tube, and the sample gas was passed over an isopropanol/dry
ice water trap. The gas was then exposed to a liquid nitrogen cold
finger, and the gaseous species that remained were evacuated. The
purified carbon dioxide was then introduced into the optical cavity.

\subsection*{Details of the 2C-CRDS Method}

The three-mirror, traveling-wave cavity, with total nominal round
trip length of 66 cm, consists of two plano mirrors and a plano-concave
mirror with 1-m radius of curvature (LohnStar). The two plano mirrors
are glued directly onto an invar cavity spacer. The concave mirror
is housed in a piezoelectric-transducer (PZT) assembly which is attached
to the invar spacer. The laser incidence angle at the PZT mirror is
\ensuremath{\sim}1.5$\degree$. The pump, probe, and reference lasers
are continuous-wave (cw) distributed-feedback quantum cascade lasers
(QCL) (Hamamatsu in high-heat-load packages). The pump (1000 mA maximum
current) and probe (500 mA maximum current) lasers in the two-color,
cavity ring-down experiments are each driven by a battery-powered
QubeCL system from ppqSense, which provides low-noise electric current
and temperature control. These lasers are modulated at 6 MHz. Light
reflection off the cavity is measured with a HgCdTe (MCT) photodetector
(Thorlabs PDAVJ8), and the MCT signal is demodulated with a frequency
mixer (Mini-Circuit, ZRPD-1+). The resulting error signal is used
as the input to the PID servo control loop (Vescent D2-125-PL) to
achieve Pound-Drever-Hall laser frequency-locking to the cavity. The
reference QCL is driven by a current controller from Wavelength Electronics
(QCL500 Laboratory Series). The temperature of the reference QCL is
regulated with a PI servo control loop (PTC2.5K-CH, Wavelength Electronics).
The reference laser was modulated at 1.7 MHz. After a double pass
through an optical cell (10 cm, 4.5 torr N$_2$O) the transmitted
intensity is recorded on an MCT photodetector (VIGO PVI-4TE-6/PIP-DC-20M)
and the signal demodulated with a frequency mixer (Mini-Circuit, ZRPD-1+).
Using wavelength modulation spectroscopy and the $\nu_{3}$ 1 \textendash{}
0, R(16) transition of $^{15}$N$^{14}$N$^{16}$O at 2214.33886 $\pm\:0.001$
cm$^{-1}$, the reference laser is locked using a PI servo loop (New
Focus LB1005).\cite{Gordon2022} The beatnote of the reference and
pump lasers is recorded on an MCT detector (VIGO PVI-4TE-10.6/FIP-1k-1G)
and provides frequency calibrations for the pump and probe lasers.
The probe laser frequency is roughly measured using a wavemeter (Bristol
771) and then assigned a frequency using the beatnote and the cavity
mode spacing. Timing for the experiment is provided with custom code
implemented on a field programable gate array (NI PXIe-7976R, 5783).
This system controls the AOMs (IntraAction Corp), provides corrections
to the PDH servo integrators, and achieves ring-down rates greater
than 2kHz. 

For the ``$^{14}$C standards'' spectra, data acquisition was started
with the cavity stabilized to zero probe detuning where ring downs
were captured for 1000 seconds. To take the spectra, an additional
15 steps were measured for 100 seconds each. The cavity PZT was stepped
using a feed-forward, pzt-creep-control profile in combination with
closed-loop control from the frequency reference.\cite{McCarttIOP2014} 

\section*{Supplementary Text}

\subsection*{Allan Deviation Analysis}

\begin{figure}[H]
\begin{centering}
\includegraphics[width=13cm]{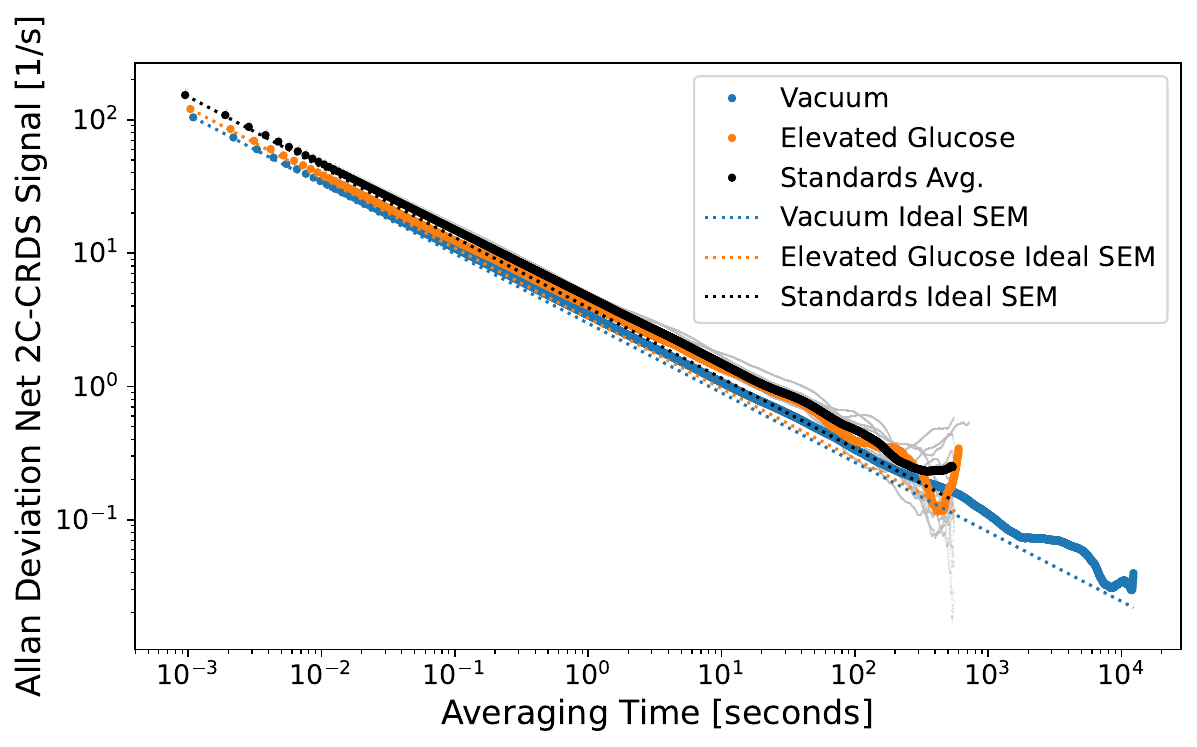}
\par\end{centering}
\caption{Allan deviation comparison of averaging stability for vacuum (blue),
$^{14}$C ``standards'' (grey and their average black), and a CO$_{2}$
sample with 375 times the natural $^{14}$C abundance (orange). Ideal
averaging precision represented by the standard error of the mean
(SEM) is indicated with dashed lines.\label{fig:Allan}}
\end{figure}

In this paper, precision values represent the stability of the 2C-CRDS
signal during the measurement of a single sample. Allan deviation
analysis allows visualization and determination of this parameter.\cite{Allan1966,Riley2008}
Figure \ref{fig:Allan} shows the overlapping Allan deviation plots
for the 2C-CRDS signal when measuring an empty cavity under vacuum,
the ``$^{14}$C standards'', and an elevated glucose sample containing
375 times the natural abundance of $^{14}$C. The fine grey lines
are the Allan deviation analysis for each of the $^{14}$CO$_{2}$
``standard'' samples from Figure 3. The black points are the average
of these $^{14}$CO$_{2}$ ``standard'' samples' individual Allan
deviations calculated from the square root of the mean of the overlapping
Allan variances. The orange points show the Allan deviation of the
elevated glucose sample. We used a sample with the largest $^{14}$CO$_{2}$
content to maximize the potential drift in the $^{14}$CO$_{2}$ 2C-CRDS
signal. Finally, the blue data points are the Allan deviation of an
empty cavity under vacuum. This is a measure of the maximum obtainable
precision of the system for a given averaging time. 

Dips in the Allan deviation below the \emph{standard error of the
mean }trend (\emph{i.e.}, $\nicefrac{\sigma}{\sqrt{N}})$ represented
by the dashed lines are likely non-physical representing a reversal
of measurement drift and insufficient averaging time. For this reason,
dips below the standard error of the mean trend seen in Figure \ref{fig:Allan}
are corrected by taking the value at the intersection of the standard
error of the mean\emph{ }trend and the Allan deviation turn around.
This is a small and conservative correction. For example, this changes
the reported elevated $^{14}$CO$_{2}$ precision from 2.6\% to 2.8\%
of the natural $^{14}$C abundance. 

\subsection*{Averaging Performance Analysis}

\begin{figure}[H]
\begin{centering}
\includegraphics[width=15cm]{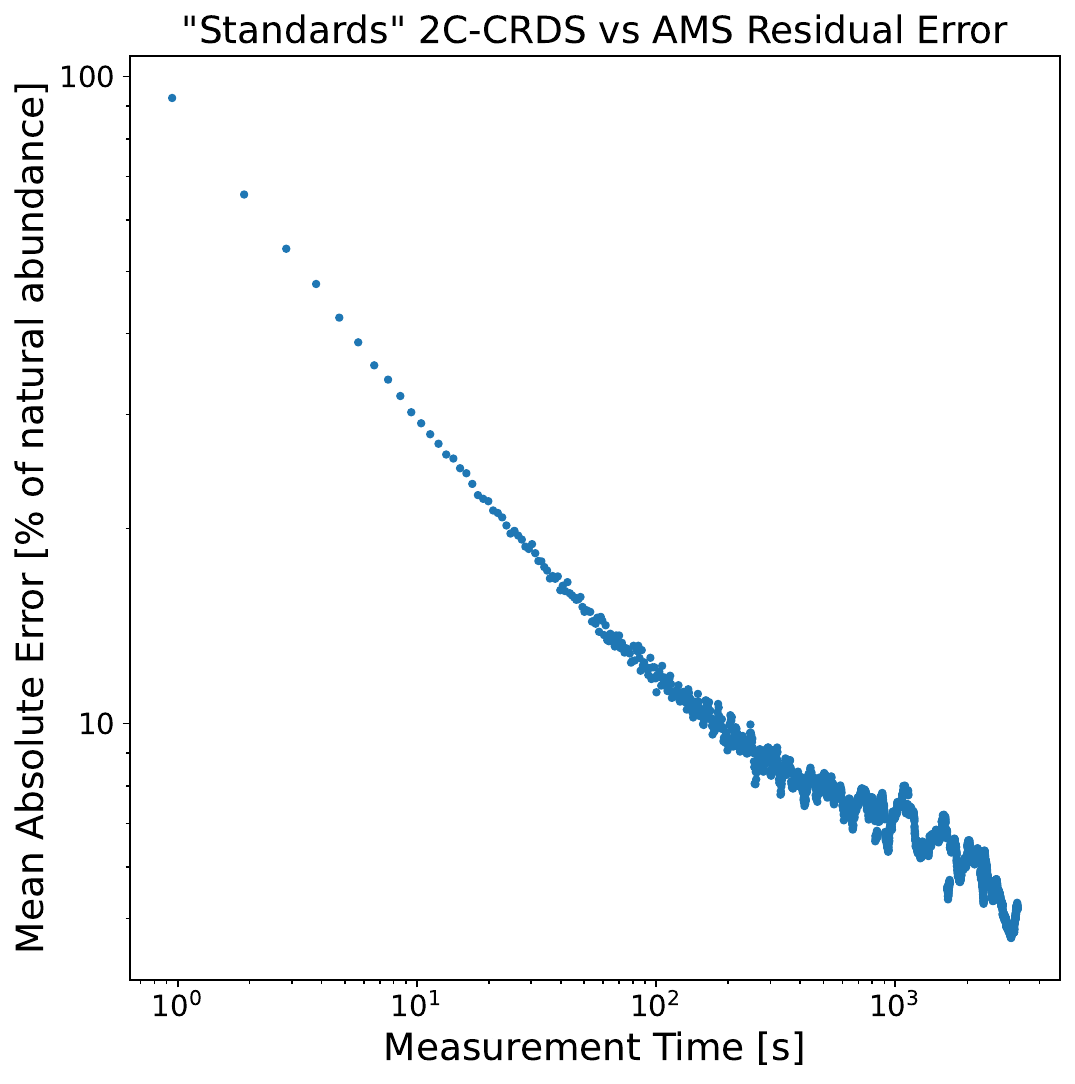}
\par\end{centering}
\caption{Analysis of ``Standards'' samples error reduction with averaging
time.\label{fig:AveragingPerformance}}
\end{figure}

Accuracy was determined by comparing 2C-CRDS and AMS measurements
using a set of ``$^{14}$C standards'' (see the main manuscript).
The residuals in Figure 3b have a mean absolute error (MAE) of 8\%.
Averaging over each sample type decreases this error to 5\% by effectively
increasing the measurement time to \textasciitilde 1 hr. The rate
of error reduction vs averaging time is visualized in Figure \ref{fig:AveragingPerformance}.
The two-color signal for each sample type is concatenated and divided
into averaging windows of the plotted measurement time. The residuals
are then calculated using the linear fit from Figure 3b.  

\subsection*{Two-Color Signal Pump Power Dependance}

\begin{figure}[H]
\begin{centering}
\includegraphics[width=13cm]{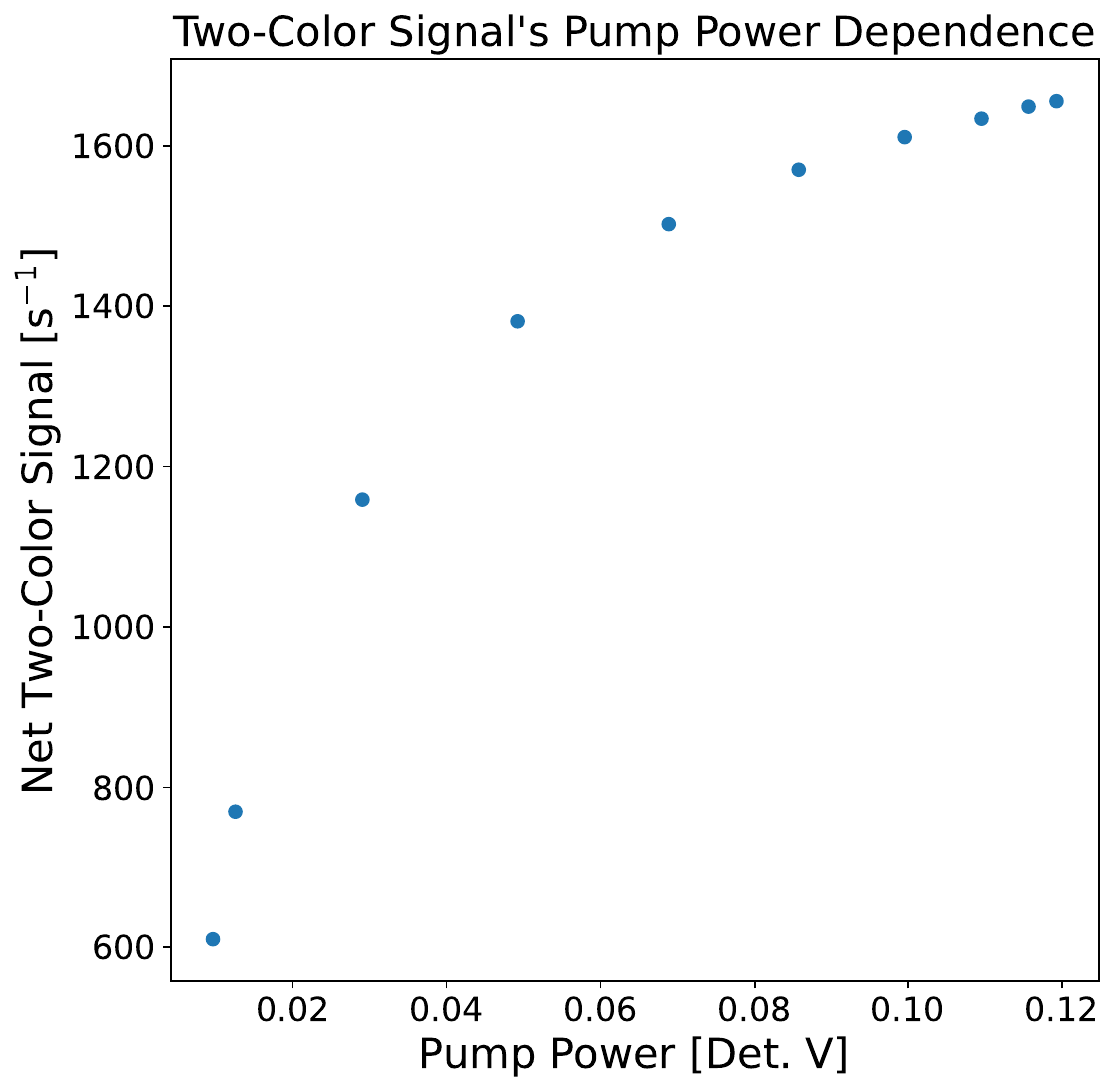}
\par\end{centering}
\caption{Pump power dependence of the net two-color signal for a 22 torr, carbon
dioxide sample with 375 time the natural abundance of $^{14}$C. The
net two-color signal is plotted versus detector voltage for the transmitted
cavity pump beam.\label{fig:power-dependence-two-color}}
\end{figure}

Figure \ref{fig:power-dependence-two-color} shows the pump-power
dependance of the net two-color signal for a 22-torr carbon-dioxide
sample with 375 times the natural abundance of $^{14}$C. The net-two
color signal is plotted vs the pump intensity transmitted through
the cavity. For this sample, the net two-color signal primarily originates
from $^{14}$CO$_{2}$ (>98\%), and the signal is well saturated with
half the applied intra-cavity pump power only reducing the net two
color signal by 12\%. 

\bibliography{citations_fixSubp_v4.bib}